\documentclass[final,4p]{elsarticle}
\usepackage{amssymb}
\usepackage{graphicx}
\usepackage{multirow}
\usepackage{txfonts}
\usepackage{doublespace}

\begin{document}

\begin{frontmatter}

\title{\bf Study of the chemical evolution and spectral signatures of some interstellar precursor molecules 
of adenine, glycine \& alanine}
\author[ICSP]{Liton Majumdar}
\ead{liton@csp.res.in}
\author[ICSP]{Ankan Das}
\ead{ankan.das@gmail.com}
\author[SNBNCBS,ICSP]{Sandip K. Chakrabarti}
\ead{chakraba@bose.res.in}
\author[MMCC,ICSP]{Sonali Chakrabarti}
\ead{sonali@csp.res.in}
\address[ICSP]{Indian Centre For Space Physics, 43 Chalantika, Garia Station Road, Kolkata 700084, India}
\address[SNBNCBS]{S.N. Bose National Center for Basic Sciences, JD-Block, Salt Lake, Kolkata,700098, India}
\address[MMCC]{Maharaja Manindra Chandra College, 20 Ramakanto Bose Street, Kolkata, 700003,India}

%\date{Accepted; Received }

\begin{abstract}
We carry out a quantum chemical calculation to obtain the infrared and electronic absorption spectra of several 
complex molecules of the interstellar medium (ISM). These molecules are the precursors of adenine, glycine \& alanine.
They could be produced in the gas phase as well as in the ice phase. We carried out a hydro-chemical 
simulation to predict the abundances of these species in the gas as well as in the ice phase. 
Gas and grains are assumed to be interacting through the accretion of various species from the gas phase on to the grain surface and 
desorption (thermal evaporation and photo-evaporation) from the grain surface to the gas phase. 
Depending on the physical properties of the cloud, the calculated abundances varies. 
The influence of ice on vibrational frequencies of different pre-biotic molecules was obtained 
using Polarizable Continuum Model (PCM) model with the integral equation formalism variant (IEFPCM) as
default SCRF method with a dielectric constant of $78.5$. Time dependent density functional theory (TDDFT) is
used to study the electronic absorption spectrum of complex molecules which are biologically important
such as, formamide and precursors of adenine, alanine and glycine.  
We notice a significant difference between the spectra of the gas and ice phase (water ice).
The ice could be mixed instead of simple water ice. We have varied the ice composition to 
find out the effects of solvent on the spectrum. We expect that our study could set the 
guidelines for observing the precursor of some bio-molecules in the interstellar space.
\end{abstract}


\end{frontmatter}

\section{Introduction}
Presence of interstellar dust towards the formation of complex interstellar molecules is taken for granted
especially after the discovery of more than twenty molecules around the star 
forming region in the interstellar ice. Altogether,
according to the CDMS catalog (http://www.astro.uni-koeln.de/cdms/molecules),
one hundred and seventy molecules have been detected in the interstellar medium or circumstellar shells.
To model the formation of complex molecules, and especially through interstellar grain chemistry, several
attempts were made by various workers (Chakrabarti et al., 2006ab; Das et al., 2008b; Das, 
Acharyya \& Chakrabarti 2010; Cuppen et al., 2007; Cuppen et al., 2009 etc.) over the years. 
Interstellar dust grains are thought to be consisting of amorphous silicate or 
carbonaceous core surrounded by molecular ice layer (Draine et al., 2003; Gibb et al., 2004). 
It has been clear from the experimental and observational results that almost 90$\%$ 
of the grain mantle is covered with H$_2$O, CH$_3$OH, CO$_2$ 
(Keane et al., 2001, Das et al., 2011 and references therein). 
Presence  of HCN, CN, CS and H$_2$O in space were identified by anomalous absorption
(Omont 1993 \& Bujarrabal et al., 1994). But a complete understanding of the chemical and physical processes 
which take place on a grain surface is still missing.

The origin of amino acids through the pre-biotic chemistry of the early earth has been a 
topic of long standing interest. However, complex pre-biotic molecules might also be formed due to very
complex and rich chemical processes inside a molecular cloud. The production of amino acids,
nucleobases, carbohydrates and other basic compounds can possibly start from the molecules 
like HCN, cyno compounds, aldehyde, and ketones (Orgel 2004; Abelson 1966), which could lead to 
the origin of life in the primitive earth conditions. However, even with the present observational 
tools, it is hard to confirm the presence of any bio-molecules in the ISM. So it may suffice, 
if we can identify a few precursor molecules which eventually form bio-molecules in the 
interstellar space. Quantum chemical simulations could be used to find out the spectral properties of  
these complex molecules. It is observed and experimentally verified that
the spectral signature of a species significantly deviates between the gas phase and the ice phase.
So a theoretical study of the spectral properties of the precursors of some important bio-molecules 
in both the gas and ice phases could serve as benchmarks for the observations. 

In this Paper, we consider a large gas-grain network coupled with a hydrodynamic simulation 
to obtain the abundances of various complex molecules, which could lead to the
formations of adenine, alanine \& glycine. We also discuss the production of formamide which is an 
important precursor in the process of the abiotic synthesis of amino acids. 
In the literature, there are several 
observational studies on glycine (Kuan et al., 2003, Hollis et al. 2003, Snyder et al. 2005, etc.). 
But its existence in a molecular cloud, till date, is not verified without a reasonable doubt. 
In case of adenine, we find that though its abundance in our theoretical model is  well under 
the observation limit, its precursor molecules are heavily abundant. It is also true for the 
alanine and glycine. These prompted us to find out the spectral signatures of the precursor 
molecules of these three molecules around the different astrophysical environment, from which 
one could roughly anticipate the abundances of adenine, glycine \& alanine. All possible reaction 
pathways are included in the gas as well as in the grain phase network. Armed with the chemical 
abundances of these precursor molecules, we compute the infrared and electronic absorption spectra 
in the gas as well as for the icy grains.  

The plan of this paper is the following. In Section 2, the models used and the computational details 
are presented. Implications of the results are discussed in Section 3. Finally, in Section 4, we draw
our conclusions.

\section{Computational details}

\subsection{Hydro-chemical Model}
The process of formation of complex molecules in the interstellar space is very much uncertain. 
There could be a number of pathways available for the formation of a complex molecule. 
However, depending on the chemical abundances of the reactive species and the reaction cross
section, the rate of formation varies. Formation routes of several interstellar bio-molecules are already 
reported in Majumdar et al., (2012). They pointed out that despite of the huge abundances of the neutral species, 
radical-molecular/radical-radical reaction pathways dominates towards the formation of 
some pre-biotic species. Normally such reactions are barrier less and exothermic in nature.
To study the chemical evolution of various complex radicals, ions, molecules which are very much 
important for the prebiotic synthesis of different bases of amino acids, we have
constructed a hydro-chemical model to mimic the interstellar scenario. 

The evolution of the chemical species is strongly dependent on the physical properties of the medium.
So the dynamic nature of the medium at any particular instant could influence 
the chemical composition of the medium. 
Das et al., (2008b) \& Das et al., (2010) considered a spherically symmetric 
isothermal (10K) collapsing cloud, whose outer boundary was assumed to be located at 
one parsec and the inner boundary was assume to be located at 10$^{-4}$ parsec. 
They used a finite difference Eulerian scheme (upwind scheme) to solve the 
Eulerian equations of hydrodynamics in spherical polar coordinates. Since they were 
interested in the spherical case, they only considered radial motion and ignored any 
dependency upon the $\theta$ \& $\phi$ coordinates. By solving the hydrodynamic equations they
studied fully time-dependent behaviour of the spherical flow.

To have a realistic condition, we have considered this density distribution as an input
for our chemical model. The gas phase chemical network is mainly adopted from the 
UMIST 2006 database (Woodall et al., 2007). 
Here, we have chosen the initial elemental abundances according 
to the Woodall et al., (2007), these are the typical low-metal abundances often 
adopted for TMC-1 cloud. We add a few new reactions following 
Chakrabarti et al., (2000ab), Woon et al., (2002), 
Quan \& Herbst (2007), Gupta et al., (2011) and references therein. 
Recently, Majumdar et al., (2012), calculated the rate coefficients 
for the reaction pathways described in Chakrabarti et al., 2000ab. They used 
Bates (1983) semi-empirical formula to find out the rate coefficients 
of any chemical reactions. Gupta et al.(2011) also followed the same prescription
to find out the reaction rates for the adenine formation in interstellar space.

To show the importance of grains towards the chemical enrichment of the ISM, 
we have also included a detailed grain chemistry network
following Hasegawa, Herbst \& Leung (1992), Das et al., (2008a), 
Das et al., (2010), Cuppen et al., (2007), Jones et al., (2011), Garrod et al., (2008) 
and Das \& Chakrabarti (2011) into our reaction network. 
We therefore have the most updated chemical network to study the 
chemical evolution of several interstellar species.
In order to perform a self-consistent study, we assume that the gas and the grains
are coupled through the accretion and the thermal evaporation processes. We
assume that the species are physisorbed onto the dust grain (classical size grain  $\sim$
1000 A$^{\circ}$) having the grain number density $1.33 \times 10^{-12}$n, 
where $n$ is the concentration of H nuclei in all forms. Thus, in principle, we have a 
complete interstellar model, which could be used to follow the hydro-chemical properties 
of a collapsing cloud. 

\subsection{Quantum chemical calculation}
First of all, we have optimized the geometry of the molecules, 
which are the precursors of various bio-molecules in space. 
In order to have an idea for the stability of these molecules,
B3LYP/6-311++G** level is used. Gas phase vibrational frequencies of these 
precursor molecules are also calculated by the B3LYP/6-311++G** level. 
{\bf Observational evidences suggest that grain mantles around the dense clouds 
are mainly covered by H$_2$O ($> 60$\%), CH$_3$OH (2-30\% with respect to solid water) 
and CO$_2$ (2-20 \% with respect to solid water). To find out the
effects of the solvent on the spectrum, we have chosen three types of ice.
(i) Unless otherwise stated, we use pure water ice. (ii) We use methanol ice also to 
mimic the ice composition around the methanol rich environment and 
finally, (iii) Based on the observational results, we construct an ice, which consists of
70\% water 20\% methanol and 10\% carbon-di oxide and call is as the `mixed ice'. }

In order to find out the vibrational frequencies of these molecules in the ice phase, we have optimized 
the geometry of these molecules in ice at B3LYP/6-311++G** level. Here, the Polarizable Continuum Model (PCM) model 
is used with the integral equation formalism variant (IEFPCM) as the default SCRF method. We have chosen 
IEFPCM model as a convenient one, since the second energy derivative is available for      
this model and also it is analytic in nature. Vibrational frequencies given here are not
exactly for the ice phase since the dielectric constant of ice ($85.5$) 
is slightly higher than that of water ($78.5$). We have calculated also
the electronic absorption spectrum of these molecules using 
the time dependent density functional theory (TDDFT study).

\section{Result and Discussion}

Till date, due to the constraints on the observational sensitivity, it is quite challenging 
to directly identify interstellar bio-molecules. For instance, the observational report 
on glycine by Kuan et al., (2003) was not 
supported by Hollis et al. (2003) and Snyder et al. (2005). Chemical models (Chakrabarti \& Chakrabarti 
2000ab, Das et al., 2008b, Majumdar et al., 2012) predict that the trace amount of
bio-molecules could be produced during the collapsing phase of a proto-star. 
Since the abundances of these molecules 
are very low, it is possible that they are not directly observable with the present day technology. 
However, if we concentrate on the pathways through which these molecules form
in the ISM and identify the precursor molecules, it could be much easier to predict their 
abundances. This is what is done in our work. We compute their chemical abundances after considering  
the gas-grain interaction in our chemical model and present the spectral signatures 
of the precursor molecules in the gas phase as well as in the grain phase.
Water is found to be the most abundant molecule followed by Methanol and Carbon-di-oxide
in the ice phase. We have concentrated on the changes of the spectral 
signature between the gas phase and ice phase.
The spectral changes with the changes of solvents are also highlighted.

\subsection{\bf Precursor molecules of adenine}
%%%%%%%%%%%%%%%%%%%%%%%%%%%%%%%%%%%%%%%%%%%%%%%%%%%%%%%%%%%%%%%%%%%%%%%%%%%%%%
\begin{figure}
\includegraphics[height=10cm,width=10cm]{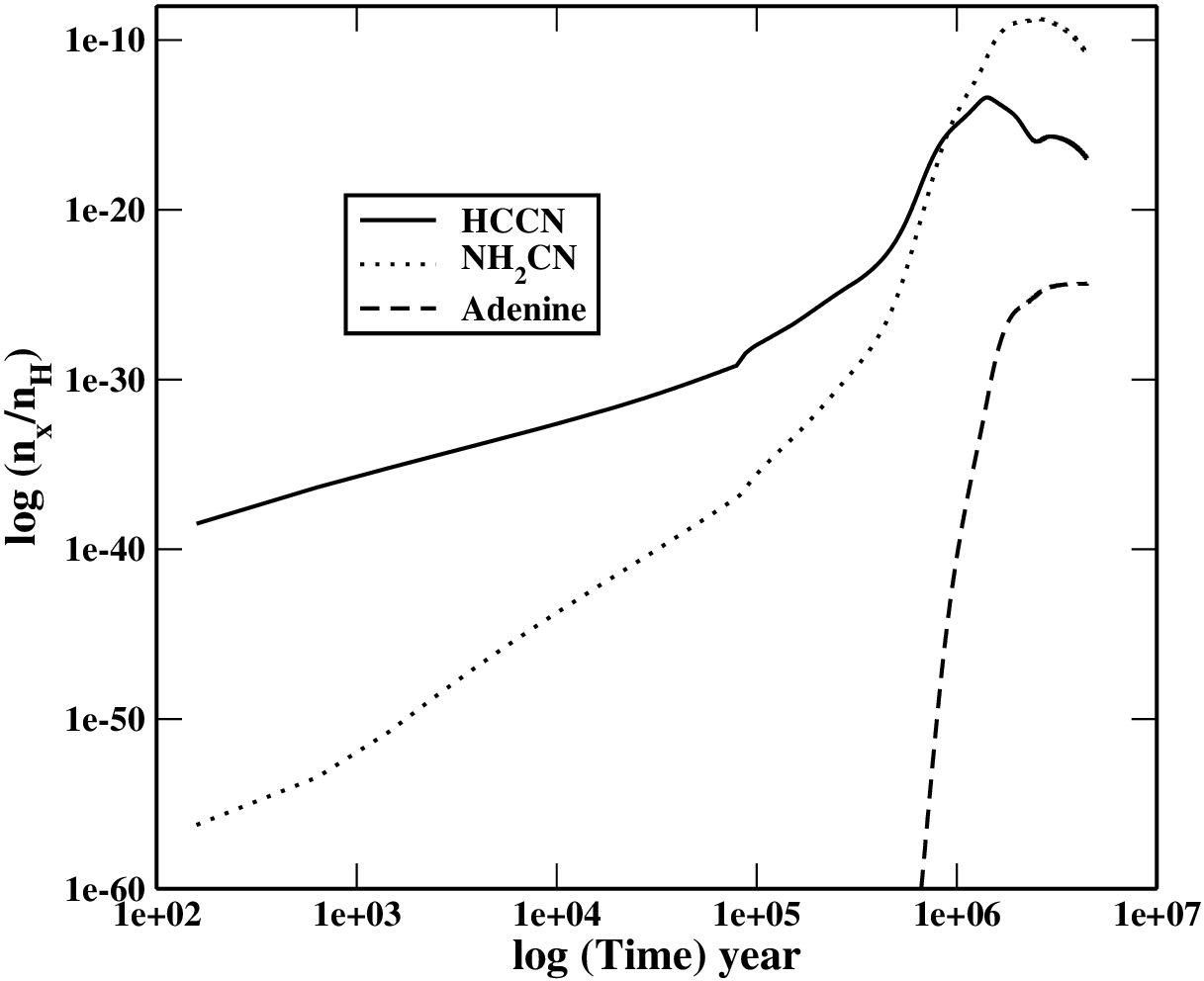}
\caption{Time evolution of adenine with its two precursor molecules. Since the precursors are several orders of magnitude higher, the probability of their detection is higher.}
\label{fig-1}
\end{figure}
%%%%%%%%%%%%%%%%%%%%%%%%%%%%%%%%%%%%%%%%%%%%%%%%%%%%%%%%%%%%%%%%%%%%%%%%%%%%%%%
There are a few studies related to the formation of adenine. Chakrabarti et al., 
(2000ab) proposed a neutral-neutral pathways for the formation of adenine. According to them, the adenine
could be produced by the following reactions:
\begin{equation}
HCN+HCN \rightarrow CH(NH)CN
\end{equation}
\begin{equation}
CH(NH)CN+HCN \rightarrow NH_2 CH(CN)_2
\end{equation}
\begin{equation}
NH_2CH(CN)_2 + HCN \rightarrow NH_2(CN)C=C(CN)NH_2
\end{equation}
\begin{equation}
NH_2(CN)C=C(CN)NH_2+HCN \rightarrow Adenine
\end{equation}
Recently Gupta et al. (2011) proposed that the following radical-molecular reaction network,
which could lead to the adenine formation.{\bf
\begin{equation}
HCCN+HCN \rightarrow C_3H_2N_2\;(1,2\; dihydro\; imidazole)
\end{equation}
\begin{equation}
C_3H_2N_2 +H \rightarrow C_3H_3N_2\;(2,3\; dihydro-1H-imidazole)
\end{equation}
\begin{equation}
C_3H_3N_2+NH_2CN \rightarrow C_4H_5N_4\;(4\; carboxaimidine-1H-imidazole)
\end{equation}
\begin{equation}
C_4H_5N_4+CN \rightarrow C_5H_5N_5 \; (2,4\; dihydro-3H-purine-6-amine) 
\end{equation}
\begin{equation}
C_5H_5N_5+H \rightarrow C_5H_6N_5\;(6\; amino-3H-purine) 
\end{equation}
\begin{equation}
C_5H_5N_5 + CN \rightarrow Adenine +HNC
\end{equation}
\begin{equation}
C_5H_5N_5+CN \rightarrow Adenine +HCN
\end{equation}}
In the chemical model described in Majumdar et al. (2012), the reactions 1 to 11 were included and  it was concluded 
that the production of adenine is dominated by the radical-molecular reaction pathways.
HCCN and NH$_2$CN are the two important 
molecules, which are responsible for the production of adenine by the radical-molecular pathways. 
So we could treat HCCN and NH$_2$CN as the precursor molecules for the production of interstellar Adenine.
HCCN is highly abundant in the interstellar space (Jiurys et al., 2006, Guelin \& Cernicharo 1991).
The formation of HCCN on the grain was considered by Hasegawa, 
Herbst \& Leung (1992). McGonagle \& Irvine (1996) 
conducted a deep search for HCCN towards TMC-1 and several GMC's via its N(J)=1(2)$\rightarrow$0(1) transition. They
set an upper limit of fractional abundance with respect to molecular hydrogen of 2 $\times 10^{-10}$.
The existence of NH$_2$CN in the interstellar cloud was first reported by Turner et al. (1975). After 
that it was observed in both diffuse and dense clouds by Liszt \& Lucas (2001). Woodall (2007) predicted a
steady state fractional abundance  of 2.02$\times 10^{-10}$ for NH$_2$CN with respect to H$_2$.

%%%%%%%%%%%%%%%%%%%%%% IR NH2CN %%%%%%%%%%%%%%%%%%%%%%%%%%%%%%%%%%%%%%%%%%%%%%%
\begin {figure}
\includegraphics[height=10cm,width=10cm]{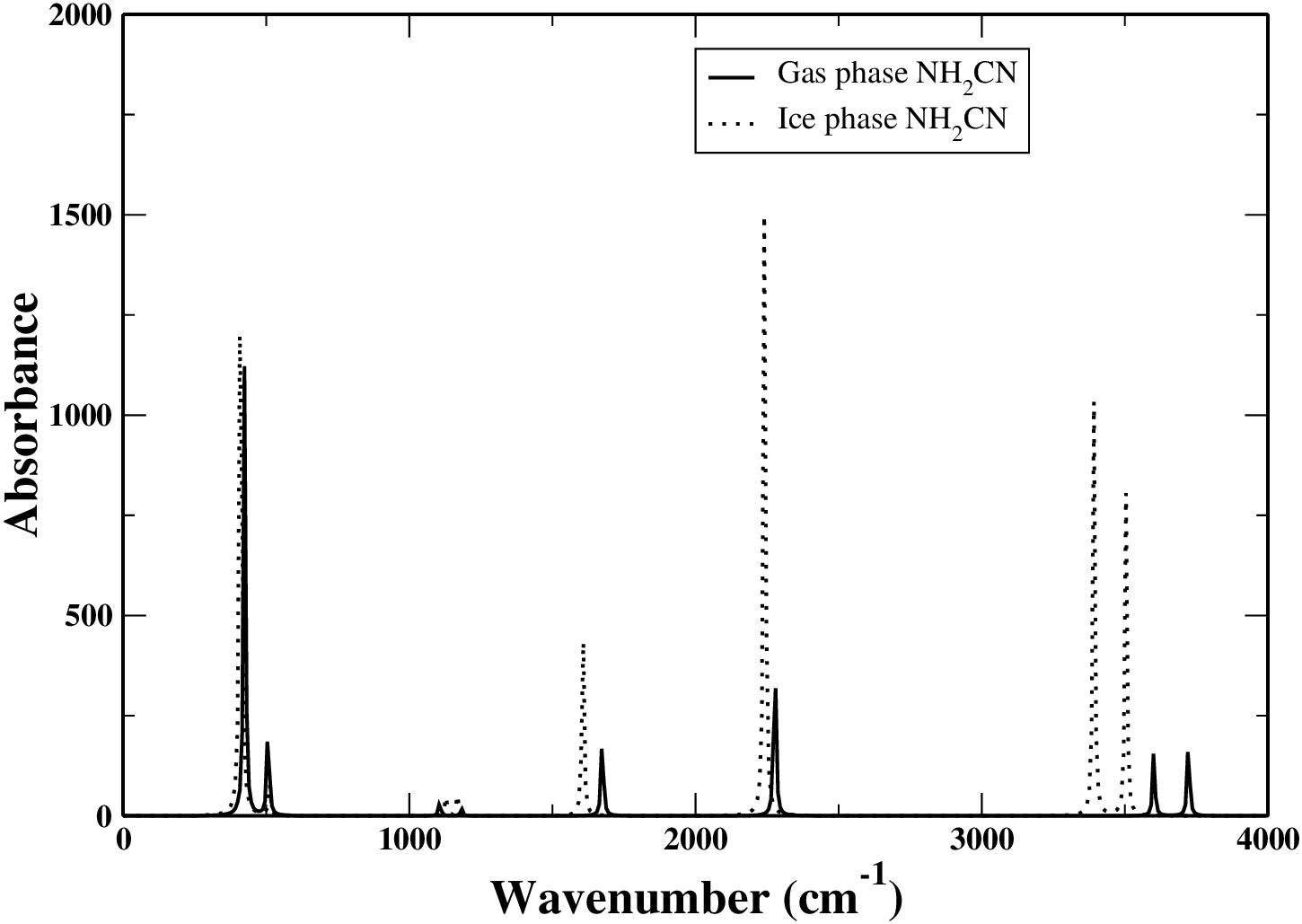}
\caption{Infrared spectrum of NH$_2$CN in gas as well as in H$_2$O ice. The line strength in the ice
phase is generally several times higher.}
\label{fig-2}
\end {figure}
%%%%%%%%%%%%%%%%%%%%%%%%%%%%%%%%%%%%%%%%%%%%%%%%%%%%%%%%%%%%%%%%%%%%%%%%%%%%%%%
We already  mentioned that we are using the density distribution of our hydrodynamical model
as an input of our chemical model. Since production of complex species in the 
intermediate region of the cloud is favourable, we consider the density distribution of intermediate
region as an input of our chemical model. 
In Fig. 1, we present the chemical evolution of adenine in gas phase along with its two precursor 
molecules, namely, HCCN \& NH$_2$CN.
We find that the abundance of NH$_2$CN is significantly higher, having a peak abundance of 1.7 $\times 10^{-9}$
and a final abundance of NH$_2$CN (after $\sim$ 5 $\times 10^6$ years) is 1.38$\times 10^{-11}$ with respect to 
total hydrogen nuclei.  
{\bf In our case, final abundance of NH$_2$CN appears to be slightly lower than the amount 
calculated by Woodall et al., (2007). This is probably because Woodall et al., (2007) 
carried out their simulation by assuming a constant density 
cloud (10$^4$ cm$^{-3}$), whereas in our case, we consider a realistic
density distribution obtained from the isothermal hydrodynamic model. 
Moreover, our chemical network consists of a large gas-grain network 
whereas the grain-surface network is missing in Woodall et al., (2007).}
Peak value of the HCCN abundance is found to be 4.1 $\times 10^{-14}$ and finally it turns out to be 
1.14$\times 10^{-17}$. With these two significantly abundant molecules, 
adenine could be produced with an abundance of 4.4 $\times 10^{-25}$.
In our hydrodynamic model, density around the inner grid locations
started to increase rapidly after 10$^5$ years. 
The reason behind this is that after one dynamical time scale ($\sim$ free fall time $\sim$
10$^5$ years), there are the smooth transport of matter from the outer envelope to the inner envelope.
After some dynamical times (i.e., around 10$^6$ years), core mass gets heavier and 
induce more matter from the
cloud than the matter actually accreting to the cloud through the outer envelope. Finally, 
the cloud relaxes to a steady state density distribution.
Since there was a sharp rise in density between 10$^5$-10$^6$ years for our
hydrodynamical model, our chemical evolution also shows (Fig. 1) sharp rise in 
the abundances during that period.
Since we chose the initial elemental abundances following Woodall et al. (2007),
basically, we started with zero abundance of any complex species. 

NH$_2$CN is a planar pentatonic molecule. It has nine fundamental vibrations (Rubalcava, 1956).
In Fig. 2, infrared absorption spectra of NH$_2$CN for the gas phase and ice phase is shown. 
We note that the gas phase NH$_2$CN has its strongest feature at
424 cm$^{-1}$ and the second intense peak arises at 2277 cm$^{-1}$. 
In the ice phase 424 cm$^{-1}$ peak
is shifted to the left and appear at 411 cm$^{-1}$ and peak at 2277 cm$^{-1}$
is also shifted in the left and appear at 2241 cm$^{-1}$ with much higher intensity.
Rubalcava (1956) assigned a vibrational peak for the gas phase 
NH$_2$CN at 2275 cm$^{-1}$ which was shifted at 2257 cm$^{-1}$
in the solution of methylene chloride at 25$^\circ$C.
Beside these intense peaks there are several peaks which are pronounced 
in our ice phase model (such as peaks at
3391 cm$^{-1}$ and 3504 cm$^{-1}$). For the sake of better understanding, we
tabulate all the infrared peak positions with their absorbance in Table 1.
%%%%%%%%%%%%%%%%%%%%%%%%%%%%%%%%%%%%%%%%%%%%%%%%%%%%%%%%%%%%%%%%%%%%%%%%%%%%%%%%%%%%%%%%%%%%%%%%%%%%%%%%%%%%
\begin{table*}
\scriptsize{
\centering
\hskip -2.5cm
\vbox{
%\addtolength{\tabcolsep}{-6pt}
\caption{Vibrational frequencies of different complex molecules in gas phase, H$_2$O ice and methanol containing grains at B3LYP/6-311G++** level}
\begin{tabular}{|c|c|c|c|c|c|c|}
\hline
{\bf Species}&{\bf Peak positions }&{\bf Absorbance}&{\bf Peak positions }&{\bf Absorbance}&{\bf Peak positions }&{\bf Absorbance}\\
{}&{\bf (Gas phase)}&{}&{\bf (H$_2$O ice)}&{}&{\bf (Methanol ice) }&{}\\
&\bf (Wavenumber in cm$^{-1}$)&&\bf (Wavenumber in cm$^{-1}$)&&\bf (Wavenumber in cm$^{-1}$)&\\
\hline
&361.81&0.02&365.76&.001&365.66&0.002\\
&472.51&0.48&476.23&1.84&476.15&1.78\\
{\bf HCCN}&953.94&2.43&962.29&3.9&962.2&3.85\\
&1264.15&3.11&1261.21&20.78&1261.32&19.79\\
&1837.77&1.67&1842.18&.33&1842&0.36\\
&3442.74&72.88&3256.78&296.99&3264.26&285.76\\
\hline
&399.44&1.18&411.2&558.25&411.91&549.69\\
&424.58&331.27&419.28&0.64&418.19&1.36\\
&506.79&78.52&511.60&21.81&511.20&23.53\\
&1106.5&11.24&1129.17&14.95&1128.48&14.9\\
{\bf NH$_2$CN}&1182.24&5.81&1167.57&15.23&1168.24&14.82\\
&1674.56&68.34&1606.43&144.79&1609.51&141.85\\
&2277.03&142.76&2241.40&486.82&2242.75&471.1\\
&3601.35&49.83&3391.73&300.21&3400.13&287.59\\
&3722.54&64.62&3504.14&232.97&3512.95&225.09\\
\hline
&318.67&500.63&395.86&381.58&399.5&392.74\\
&906.57&7.79&454.78&36.29&433.45&36.96\\
&1247.51&43.21&484.98&287.53&496.42&262.89\\
&1295.04&9.11&927.74&5.58&929.13&5.2\\
&1488.33&0.03&1232.40&54.84&1233.35&54.55\\
&1697.65&61.9&1305.44&11.38&1306.49&11.11\\
{\bf CH$_2$NH$_2$}&3175.48&11.04&1482.85&2.88&1484.23&2.92\\
&3302.9&13.91&1658.80&105.1&1660.7&103.69\\
&3618.37&18.34&3121.60&19.78&3120.02&19.94\\
&3744.57&20.4&3244.2&22.13&3241.7&22.41\\
&-&-&3454.87&83.77&3463.3&79.24\\
&-&-&3574.02&87.41&3582.74&83.04\\
\hline
&563.02&174.43&527.91&274.95&525.87&270.55\\
&564.06&37.16&576.96&88.35&576.21&85.09\\
{\bf COOH}&971.71&185.63&1004.45&360.38&1005.32&353.26\\
&1246.79&2.28&1238.63&12.42&1242.43&11.59\\
&1719.72&337.79&1661.63&634.66&1663.93&619.87\\
&3427.26&1.56&2759.29&339.39&2808.53&311.42\\
\hline
&93.80&73.46&227.10&24.35&219.98&24.79\\
&227.28&12.91&236.09&21.97&235.28&21.72\\
&654.85&5.11&659.67&12.28&659.02&11.85\\
&679.23&105.13&698.25&142.52&696.02&141.02\\
&933.22&9.24&932.94&1.37&932.81&1.33\\
&1133&357.05&1086.56&948.30&1086.95&923.82\\
{\bf C$_2$H$_3$ON}&1187.76&184.8&1187.07&176.63&1187.21&178.39\\
&1453.73&1.06&1443.51&4.87&1443.04&4.98\\
&1476.31&0.67&1484.93&0.19&1484.08&0.16\\
&2127.17&299.17&2131.47&734.61&2133.49&720.19\\
&3229.69&2.38&3205.66&1.28&3209.03&1.32\\
&3384.87&7.47&3365.84&11.18&3369.44&11\\
&3505.82&84.71&3474.5&180.74&3479.12&177.75\\
\hline
&191.8&3.32&191.32&15.01&194.43&10.05\\
&218.57&21.96&222.92&3.13&224.96&4.01\\
&267.93&27.05&263.17&186.82&265.86&78.55\\
&290.27&130.59&278.65&134.51&298.21&198.11\\
&387.79&19.01&392.16&18.82&394.74&19.99\\
&570.49&11.56&573.19&6.55&573.64&7.09\\
&584.91&4.19&596.31&2.59&596.93&2.57\\
&788.74&16.30&802.39&8.91&893.81&80.95\\
&902.23&31.88&894.69&79.43&1032.76&89.02\\
&1033.80&83.20&1031.17&108.92&1089.89&11.37\\
&1094.69&10.61&1085.35&12.45&1137.77&44.92\\
&1135.26&47.07&1133.81&48.79&1273.41&22.11\\
{\bf C$_3$H$_5$ON}&1278.81&10.48&1260.6&63.62&1373.41&22.11\\
&1356.76&3.7&1363.86&18.47&1390.68&5.87\\
&1400.34&23.91&1378.94&11.81&1447.1&13.89\\
&1449.81&9.51&1443.03&12.52&1509.23&13.09\\
&1523.18&10.39&1509.64&11.54&1520.34&10.94\\
&1526.22&7.22&1521.19&9.31&2272&11.98\\
&2277.03&0.9&2270.35&19.92&3042.14&13.61\\
&3027.25&10.75&2972.02&2.17&3048.99&9.08\\
&3079.31&0.37&3032.65&11.22&3122.11&22.68\\
&3104.22&22.68&3112.9&16.33&3130.03&23.59\\
&3121.81&13.86&3121.92&16.47&3659.5&46.27\\
&3646.01&20.12&3243.97&248.16&-&-\\
\hline
\end{tabular}}}
\end{table*}
In Table 1, we note down the peak position of the infrared spectrum of 
cyanocarbene (HCCN) in the gas as well as in the ice phase.  
We find that the most intense mode in the gas phase appears at 3442 cm$^{-1}$. 
This peak is shifted in the left in the ice phase by 186 cm$^{-1}$  
and appears at 3256 cm$^{-1}$. Other peaks in
the gas phase have negligible contributions. 
One weak peak at 1261 cm$^{-1}$ appears in the ice phase. 
McCarthy et al. (1995) detected rotational transitions from seven 
low-lying vibrational states of HCCN with
a sensitive millimeter-wave spectrometer. They obtained a peak at 
3256 cm$^{-1}$, which is exactly coinciding with our
strongest peak at the ice phase. Dendramis and Leroi (1977)
performed a matrix isolation spectroscopy to find out the
infrared signatures of the vibrational modes of HCCN. They obtained 
this peak at around 3229 cm$^{-1}$. 

%%%%%%%%%%%%%%%%%%%%%%%%%%%% EE NH2CN%%%%%%%%%%%%%%%%%%%%%%%%%%%%%%%%%%%%%%%%%%
\begin {figure}
\vskip 1cm
\includegraphics[height=10cm,width=10cm]{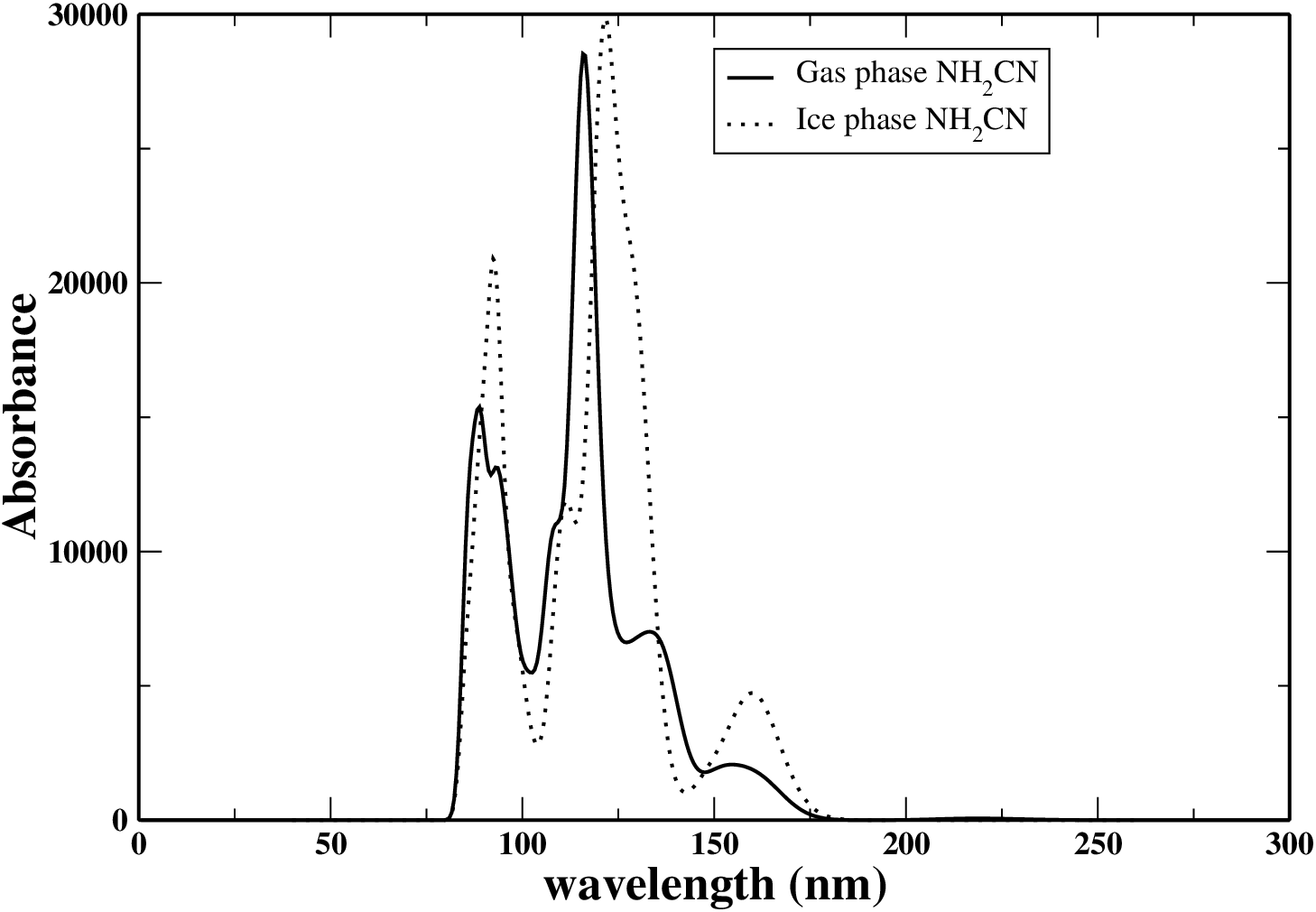}
\caption{Electronic absorption spectra of NH$_2$CN in gas as well as in H$_2$O ice}
\label{fig-3}
\end {figure}
%%%%%%%%%%%%%%%%%%%%%%%%%%%%%%%%%%%%%%%%%%%%%%%%%%%%%%%%%%%%%%%%%%%%%%%%%%%%%%%%%

%%%%%%%%%%%%%%%% Electronic Part for different precursor %%%%%%%%%%%%%%%%%%%%%%%%%%%%%%%%%%%%%%%%%%%%%%%%%%%%%%%%%%%%%%%%%%%%%%

\begin{table*}
\centering
\addtolength{\tabcolsep}{-4pt}
\scriptsize{
\hskip -2.5cm
\vbox{
\caption{Electronic transitions of different complex molecules at B3LYP/6-311++G** level theory   in gas phase and H$_2$O ice}
\begin{tabular}{|c|c|c|c|c|c|c|c|c|c|c|}
\hline
{\bf Species}&{\bf Wavelength}&{\bf Absorbance}&{\bf Oscillator strength }&{\bf Transitions}&{\bf Contribution}&{\bf Wave length}&{\bf Absorbance}&{\bf Oscillator strength}&{\bf Transitions}&{\bf Contribution}\\
{}&{(gas phase)}&{}&{}&{}&{(\%)}&{(H$_2$O ice)}&{}&{}&{}&{(\%)}\\
&(in nm)&&&&&(in nm)&&&&\\
\hline
&156.2&2058&0.0202&H-0$\rightarrow$ L+5&38&159.2&4713&0.0837&H-0$\rightarrow$ L+4&73\\
&133.7&7003&0&H-1$\rightarrow$ L+5&86&121.9&10761&0.0002&H-0$\rightarrow$ L+8&93\\
{\bf NH$_2$CN}&114.83&25998&0.0272&H-2$\rightarrow$ L+5&81&111.6&11870&0&H-0$\rightarrow$ L+12&91\\
&88.9&15378&0.1128&H-3$\rightarrow$ L+9&72&92.2&20896&0.0168&H-1$\rightarrow$ L+12&97\\
\hline
&397.4&656&0.0162&H-0$\rightarrow$ L+1&201&365.9&734&0.0176&H-0$\rightarrow$ L+1&201\\
&279.6&1560&0.0376&H-0$\rightarrow$ L+4&199&285.7&2255&0.0548&H-0$\rightarrow$ L+3&198\\
{\bf CH$_2$NH$_2$}&187.6&4774&0.0066&H-0$\rightarrow$ L+10&103&203.0&2505&0.1055&H-1$\rightarrow$ L+1&161\\
&122.0&9633&0.011&H-2$\rightarrow$ L+2&134&151.0&9357&0.0459&H-1$\rightarrow$ L+0&166\\
&104.0&6197&0.0066&H-1$\rightarrow$ L+12&151&121.2&8975&0.0009&H-1$\rightarrow$ L+4&195\\
\hline
&262.9&540&0.0120&H-1$\rightarrow$ L+1&191&259.2&738&0.0174&H-2$\rightarrow$ L+1&192\\
&219.2&2777&0.0683&H-0$\rightarrow$ L+1&194&200.2&1747&0.0363&H-0$\rightarrow$ L+1&192\\
{\bf COOH}&158.4&4349&0.0037&H-2$\rightarrow$ L+0&120&169.2&6085&0.0936&H-0$\rightarrow$ L+3&119\\
&126.4&5938&0.0002&H-3$\rightarrow$ L+0&122&124.1&11461&0.0162&H-1$\rightarrow$ L+2&130\\
&123.2&5887&0.0662&H-0$\rightarrow$ L+7&149&101.0&6861&0.0086&H-1$\rightarrow$ L+6&90\\
&114.4&5581&0.00005&H-2$\rightarrow$ L+3&86&-&-&-&-&-\\
&101.0&3678&0.0121&H-0$\rightarrow$ L+10&173&-&-&-&-&-\\
\hline
&306.8&6245&0.1448&H-0$\rightarrow$ L+2&93&309.0&8194&0.0108&H-0$\rightarrow$ L+3&98\\
{\bf C$_2$H$_3$ON}&190.0&10132&0.0002&H-0$\rightarrow$ L+9&88&204.4&4352&0.0063&H-0$\rightarrow$ L+7&88\\
&113.0&12982&.0287&H-0$\rightarrow$ L+20&96&106.3&20307.56&0.0095&H-2$\rightarrow$ L+9&71\\
\hline
&144.73&6098&0.0317&H-1$\rightarrow$ L+2&24&-&-&-&-&-\\
{\bf C$_3$H$_5$ON}&118.0&19512&0.0178&H-1$\rightarrow$ L+8&59&118.9&26299&0.0518&H-3$\rightarrow$ L+5&61\\
\hline
\end{tabular}}
}
\end{table*}
%%%%%%%%%%%%%%%%%%%%%%%%%%%%%%%%%%%%%%%%%%%%%%%%%%%%%%%%%%%%%%%%%%%%%%%%%%%%%%%%%%%%%%%%%%%%%%%%%%%%%%%%%%%%%%
We continue our computation to obtain the spectral properties of HCCN in the 
electronic absorption mode. 
However, it does not appear to have any significant contribution.
Electronic absorption spectra of NH$_2$CN molecule in gas (solid) as well 
in the ice phase (gas) are shown in Fig. 3. 
The electronic transitions, absorbance and percentage of contribution 
are summarized in Table 2. An electronic absorption spectrum of NH$_2$CN 
molecule in the gas phase is characterized by four intense peaks. These four
transitions occurred at $156$, $133$, $114$, $89$nm with major 
contributions from H-0$\rightarrow$ L+4, 
H-0$\rightarrow$ L+8, H-0$\rightarrow$ L+12, H-1$\rightarrow$ L+12 transitions 
respectively. Here, H represents 
the highest occupied molecular orbital (HOMO) and L represents the lowest 
unoccupied molecular orbital (LUMO). 
Peaks at $156$nm and $89$nm are shifted to the right in the ice phase and appear at
$159$nm and $92$nm respectively with a bit high intensities.
The peak at $133$nm is shifted to $121$nm and has a high intensity. The
peak at $114$nm is shifted to $111$nm and has a lower intensity.
All the peak locations in electronic absorption spectra with their respective
absorbance and oscillator strengths are highlighted in Table 2.

\subsection{\bf Precursor molecules of glycine}
%%%%%%%%%%%%%%%%%%%%%%%%%%%%%%%%%%%%%%%%%%%%%%%%%%%%%%%%%%%%%%%%%%%%%%%%%%%%%%
\begin{figure}
\vskip 1cm
\includegraphics[height=10cm,width=10cm]{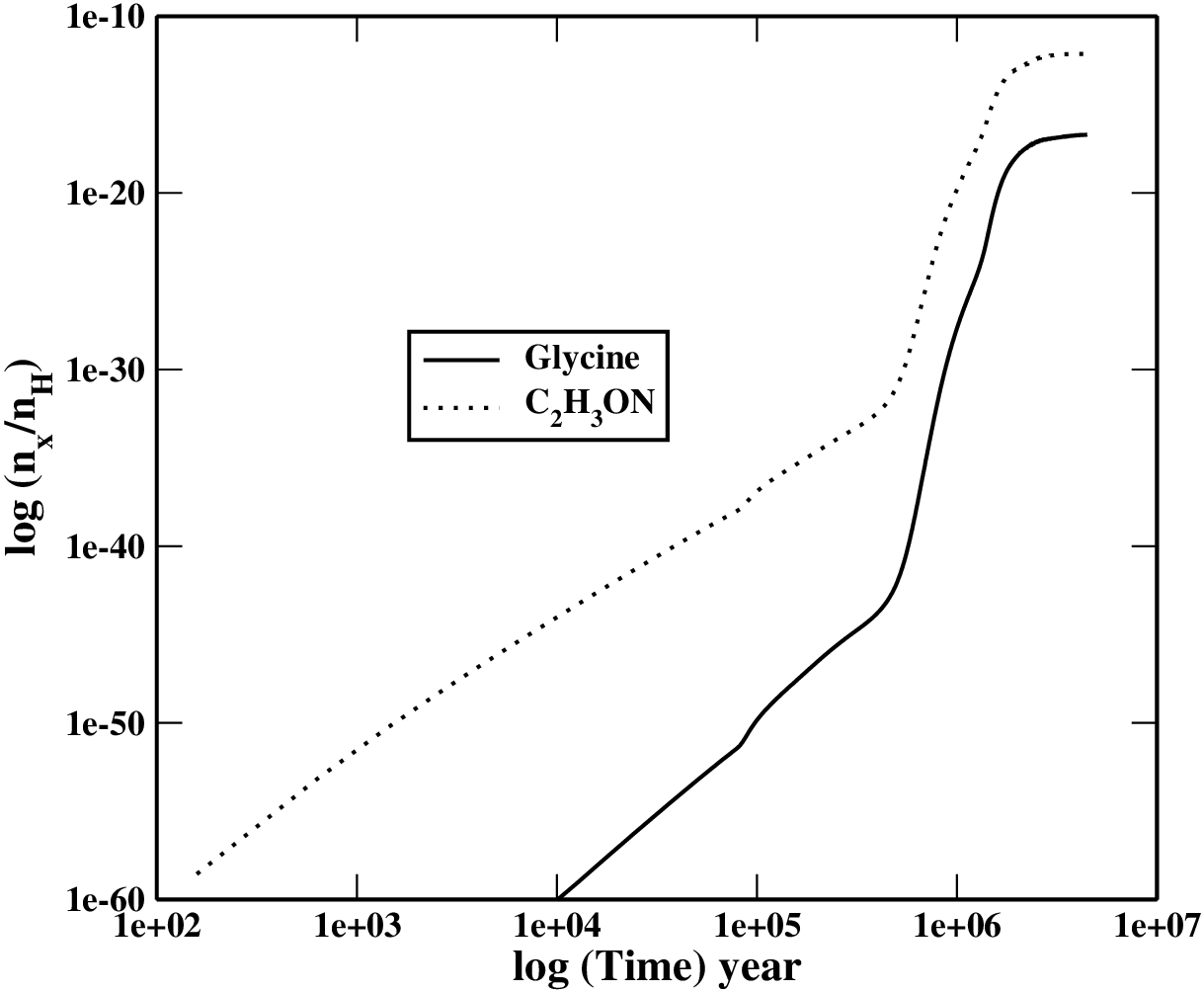}
\caption{Time evolution of glycine with its precursor molecule, C$_2$H$_3$ON. 
The abundance of the precursor 
molecule is much higher and could be observed.}
\label{fig-4}
\end{figure}
%%%%%%%%%%%%%%%%%%%%%%%%%%%%%%%%%%%%%%%%%%%%%%%%%%%%%%%%%%%%%%%%%%%%%%%%%%%%%%%
Observations of glycine are highly debated till date. Several chemical modeling as well as 
experiments were performed during the past years. Chakrabarti et al. (2000a) 
proposed the following pathways for the glycine production;
\begin{equation}
H_2 CO+HCN \rightarrow C_2H_3ON
\end{equation}
\begin{equation}
C_2H_3ON+H_2O \rightarrow C_2H_5NO_2.
\end{equation}
Woon et al., (2002) performed quantum chemical calculations 
to evaluate the viability of various pathways to the formation of glycine,
They considered following pathway:
\begin{equation}
HCN+H \rightarrow HCNH
\end{equation}
\begin{equation}
HCNH+H \rightarrow CH_2NH
\end{equation}
\begin{equation}
CH_2NH+H \rightarrow CH_2NH_2
\end{equation}
\begin{equation}
CO+OH \rightarrow COOH
\end{equation}
\begin{equation}
CH_2NH_2+COOH \rightarrow C_2H_5NO_2.
\end{equation}
According to Woon (2002) glycine formation by this method is most likely
to happen in such kind of interstellar ices which could have
experienced thermal shocks or could have formed in comets that could 
have passed through warmer regions of the solar system.
Since in our model calculations, we are restricted to the temperature 10K, this network does not
influence our glycine production at all. So, in our case, glycine is mainly produced
by the neutral-neutral pathways, as first discussed by Chakrabarti et al. (2000a).
In Fig. 4, the time evolution of glycine along with its precursor 
molecule (C$_2$H$_3$ON) is shown. Peak abundance of C$_2$H$_3$ON and glycine are
found to be 7.3$\times 10^{-13}$ and 1.96$\times 10^{-17}$ respectively. 
According to Chakrabarti et al. (2000a) and Majumdar et al. (2012), 
glycine could react with H$_2$O to form glycolic acid as the following;
\begin{equation}
C_2H_5NO_2 + H_2O \rightarrow C_2H_4O_3 + NH_3.
\end{equation}
We have included this reaction in our network. Peak abundance of 
glycolic acid is calculated to be 1.36$\times 10^{-21}$. As in Fig. 1, here also
a sharp rise in the abundances are observed due to the same reason as discussed in Fig. 1.

Though the pathways proposed by Woon (2002) which does not influence the production of glycine at this
low temperature, for the sake of completeness, we have identified CH$_2$NH$_2$ and COOH
as precursor molecules for the production of glycine by this route.
Infrared peak positions with their absorbance in the gas phase as well as in the ice phase
is pointed for the species CH$_2$NH$_2$ in Table 1.  
We find that the most intense mode in the gas phase appears nearly at 318 cm$^{-1}$. This peak
is shifted to the right in the ice phase by 77 cm$^{-1}$ 
and is appearing at 395 cm$^{-1}$. It is interesting to note that
one strong peak appears at 484 cm$^{-1}$ in the ice phase and 
its corresponding peak in the gas phase spectrum is missing. 
Several new peaks are significantly pronounced in the ice phase. 
Similarly, for COOH, we have presented similar parameters in Table 1. 
The gas phase infrared spectrum of COOH mainly consists of three strong peaks.
The strongest peak in the gas phase arises at 1719 cm$^{-1}$ 
(giving an excellent agreement with Chen et al., 1998) followed by two 
moderate sized peaks at 971 cm$^{-1}$ and 
563 cm$^{-1}$ respectively. The strongest peak is shifted to the left
in the ice phase and appears at around 1661 cm$^{-1}$
followed by one right shifted peak at around 1004 cm$^{-1}$ and one 
left shifted peak at around 527 cm$^{-1}$ respectively. 
One more prominent peak in the ice phase is pronounced at 2759 cm$^{-1}$, 
which is absent in the gas phase. All the
ice phase peaks are more intense in comparison to those of the gas phase. 
Infrared spectral parameters
of C$_2$H$_3$ON in gas phase as well as in the ice phase are noted in Table 1. 
We find that gas phase IR spectra of C$_2$H$_3$ON contains 
two major peaks, one at 1133 cm$^{-1}$ and the other at 2127.17 cm$^{-1}$. From Table 1 , 
it is clear that most of the peaks in gas phase are 
shifted towards the left for the ice phase spectrum. 

Different electronic absorption spectral parameters of CH$_2$NH$_2$ in the gas phase and
in the ice phase are given in Table 2. 
In the gas phase, the spectrum is characterized by five intense peaks at $397.4$, $279.4$, $187.6$, 
$122.0$ and $104.0$ nm. These intense peaks are assigned due to the H-0$\rightarrow$ L+1,
H-0$\rightarrow$ L+4, H-0$\rightarrow$ L+10, H-2$\rightarrow$ L+2,  H-1$\rightarrow$ L+12 HOMO-LUMO transitions. 
Peak positions are slightly shifted in the ice phase. In case of COOH also, several 
intense peaks are prominent in the gas phase at $262.9$, 
$219.2$, $158.4$, $126.4$, $123.2$, $114.4$ and $101.0$ nm. Few of these peaks disappear
in the ice phase (Table 2). It is to be noted that most of the peak intensities 
in the ice phase are considerably higher. Similarly, the electronic absorption spectra of C$_2$H$_3$ON 
in the gas and ice phases are given in Table 2. In the gas phase, it is characterized by three intense peaks at 
$306.8$, $190.0$, and $113.0$ nm. These intense peaks are assigned due to the H-0$\rightarrow$ L+2,
H-0$\rightarrow$ L+9, H-0$\rightarrow$ L+20 HOMO-LUMO transitions.
These peak positions are slightly shifted in the ice phase.

%%%%%%%%%%%%%%%%%%%%%%%%%%%%%%%%%%%%%%%%%%%%%%%%%%%%%%%%%%%%%%%%%%%%%%%%%%%%%%%%%%%%%
         
\subsection{\bf Precursor molecules of alanine}
%%%%%%%%%%%%%%%%%%%%%%%%%%%%%%%%%%%%%%%%%%%%%%%%%%%%%%%%%%%%%%%%%%%%%%%%%%%%%%
\begin{figure}
\includegraphics[height=10cm,width=10cm]{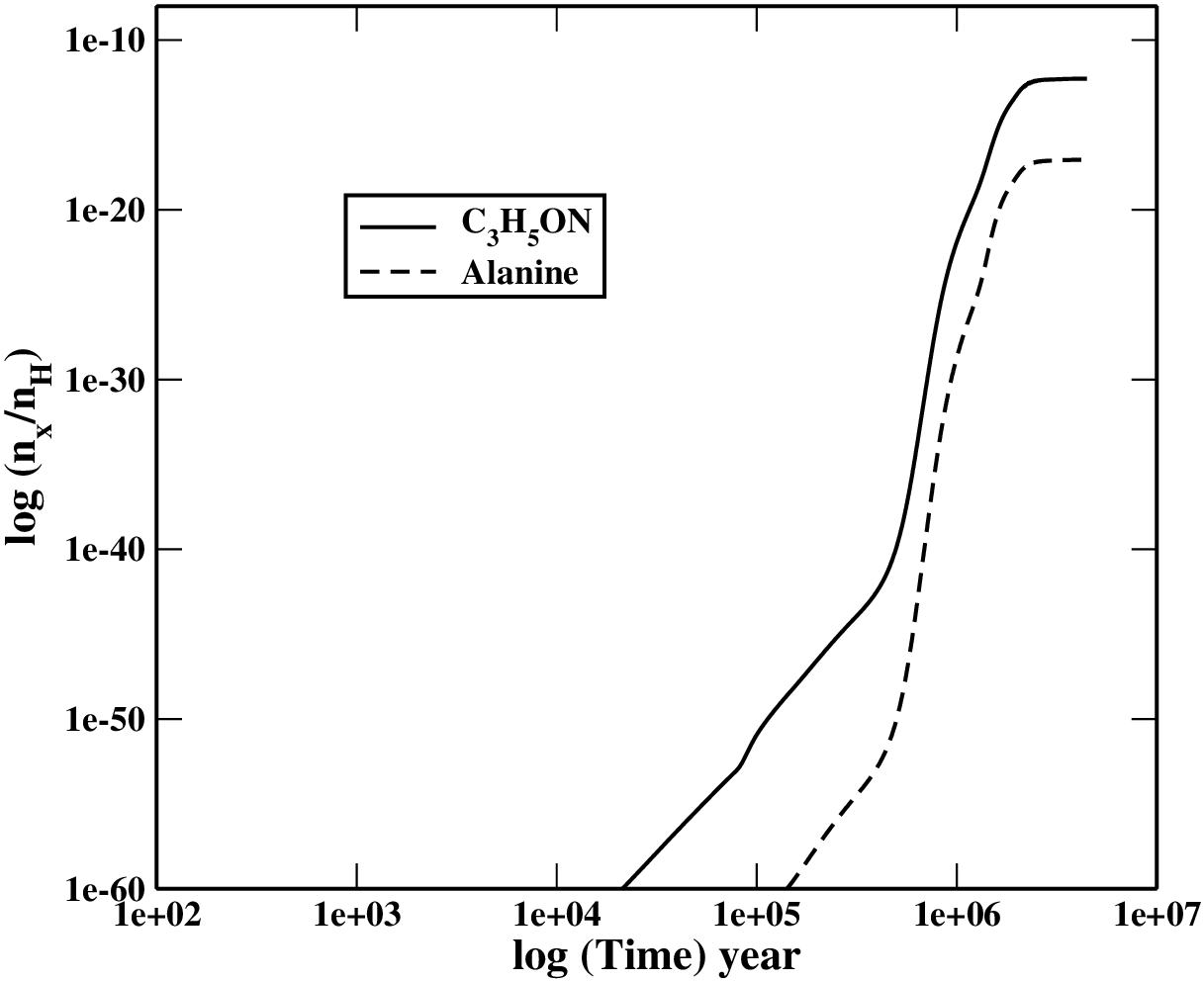}
\caption{Time evolution of alanine with its precursor molecule, C$_3$H$_5$ON.}
\label{fig-5}
\end{figure}
%%%%%%%%%%%%%%%%%%%%%%%%%%%%%%%%%%%%%%%%%%%%%%%%%%%%%%%%%%%%%%%%%%%%%%%%%%%%%%%
According to  Chakrabarti et al., 2000a, the alanine formation could be due to the following reactions:
\begin{equation}
CH_3CHO+HCN \rightarrow C_3H_5ON
\end{equation}
\begin{equation}
C_3H_5ON+H_2O \rightarrow C_3H_7NO_2.
\end{equation}
According to Woon et al., (2002) production could follow the following route;
\begin{equation}
NH_2CH_2+ COOH \rightarrow NH_2CHCOOH+H
\end{equation}
\begin{equation}
NH_2CHCOOH+CH_3 \rightarrow C_3H_7NO_2.
\end{equation}
Woon et al., (2002) discussed the production of Alanine
in the UV irradiated ice, which are much warmer. So this pathway 
is not relevant in the present situation.
So the neutral-neutral pathways as described by Chakrabarti et al. (2000a) could be
very useful. In the neutral-neutral reaction pathway, 
C$_3$H$_5$ON reacts with highly abundant gas phase H$_2$O to form alanine. 
Similar to the chemical evolution of adenine and glycine shown in Fig. 1 \& Fig. 4 respectively, 
chemical evolution of this precursor molecule along with the alanine is shown in Fig. 5.
Hydro-chemical modeling suggests that C$_3$H$_5$ON 
having a peak abundance of 5.3 $\times 10^{-13}$ could produce 
alanine with a peak abundance of 8.9 $\times 10^{-18}$.
Here too, the precursors are several orders of magnitude more abundant and 
they would be more easily detectable.	
 
Infrared peak positions along with the absorbance of C$_3$H$_5$ON in the gas as well as in the ice phase
are highlighted in Table 1. The gas phase spectrum 
consists of several intense peaks. There are two strong peaks located at
290 cm$^{-1}$ and 1033 cm$^{-1}$ respectively. In the ice phase, several new peaks appear
which have much higher intensity. The strongest peaks in the ice phase appear at 
3243 cm$^{-1}$, 278 cm$^{-1}$ and 
263 cm$^{-1}$ respectively. All other peak locations are given in Table 1.

The electronic absorption spectrum of C$_3$H$_5$ON molecule in gas 
phase is characterized by two intense peaks arising due to the H-1$\rightarrow$ L+2, H-1$\rightarrow$ L+8 HOMO-LUMO 
transitions. The ice phase electronic absorption spectrum is followed by only one peak (Table 2) at the wavelength 
118.9 nm. The peak positions along with all the details of the electronic absorption spectra are given in Table 2.

\subsection{Formamide: An important precursor in the abiotic synthesis of amino acids}
\begin {figure}
\vskip 4cm
\centering{
\includegraphics[height=10cm,width=10cm]{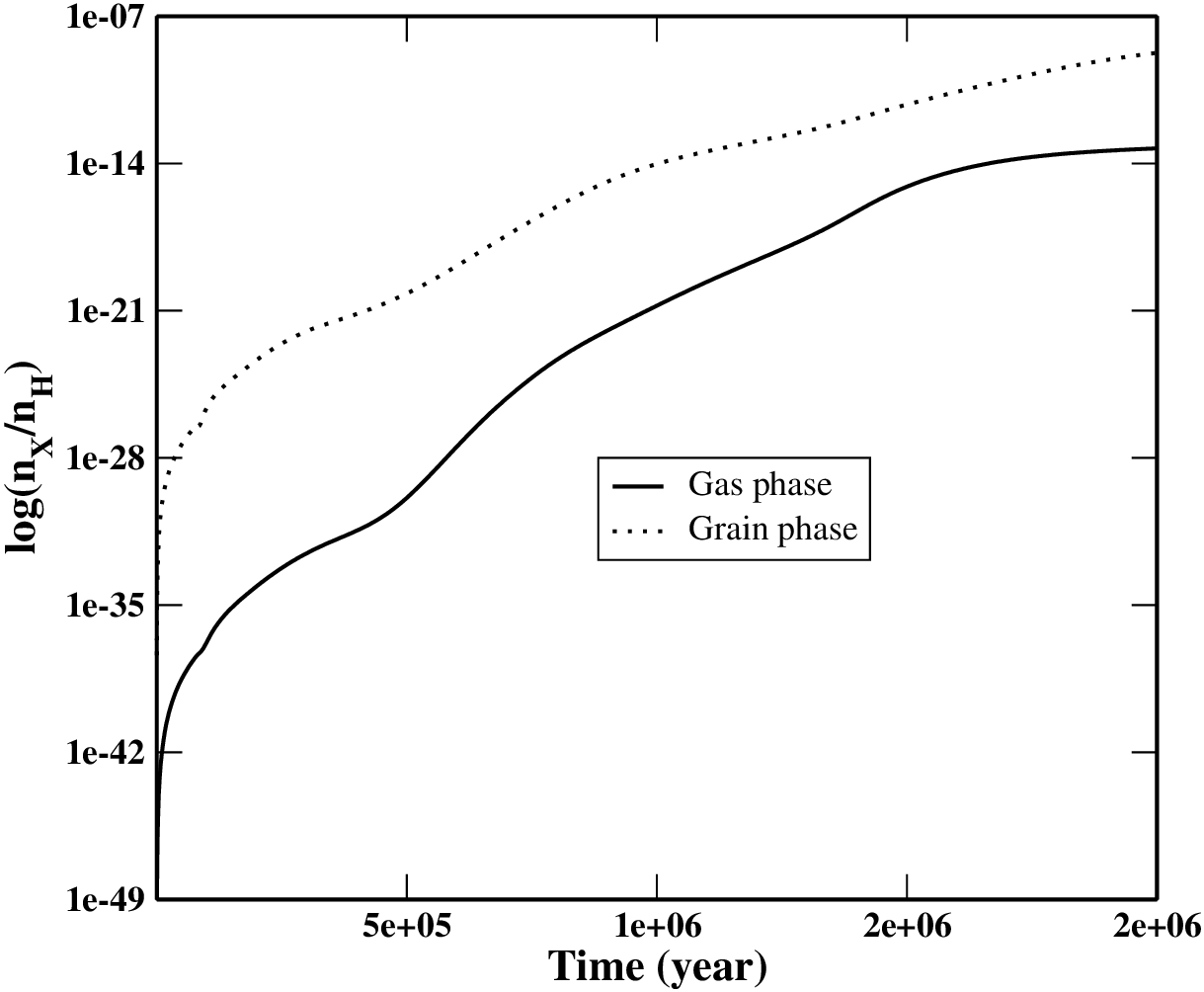}
}
\caption{Time evolution of formamide in the gas phase as well as in the ice phase.}
\label{fig-6}
\end {figure}
Formamide is the simplest amide containing peptide bond. It is very abundant 
in the ISM and could be an important precursor in the abiotic synthesis of 
amino acids and thus significant to further prebiotic chemistry in the interstellar space. 
Formamide was discovered in the interstellar space in the early 1970s. It has been identified 
by one of the gas phase molecules in the past (Millar, 2004). 
It is also highly abundant in the ice phase (Garrod et al., 2008).
Formamide in the ISM could be produced by several pathways. Here we have mainly 
followed Quan \& Herbst (2007) for its production in the gas phase. 
\begin{equation}
H_2CO + NH_4^+ \rightarrow NH_4CH_2O+ + h\nu 
\end{equation}
\begin{equation}
NH_4CH_2O +  e^- \rightarrow HCONH_2 + H + H_2.
\end{equation}
For the production of Formamide in the ice phase we follow Jones et al., (2011) who suggested 
the pathway below:
\begin{equation}
NH_3 \rightarrow NH_2+H
\end{equation}
\begin{equation}
H+CO \rightarrow HCO
\end{equation}
\begin{equation}
HCO+NH_2 \rightarrow HCONH_2,
\end{equation}
and Garrod et al., (2008) who suggested the pathway as given below:
\begin{equation}
OCN+H \rightarrow HNCO
\end{equation}
\begin{equation}
HNCO +H \rightarrow HNCHO
\end{equation}
\begin{equation}
HNCHO+ H \rightarrow HCONH_2
\end{equation}
Since reaction number $28$ could also be possible in the gas phase, we include this reaction in our
gas phase network as well. Following the same technique used in Majumdar et al., (2012), the reaction
energy for this reaction is calculated to be $-4.12$ eV and the 
rate coefficient calculated to be 2.73$\times 10^{-11}$ cm$^3 S^{-1}$.
In Fig. 6, we have shown the time evolution of the formamide in the gas phase as well as
in the ice phase.
The peak abundance of the gas phase formamide is calculated to be 1.33 $\times$ 10$^{-13}$, whereas the
grain phase formamide appears to be highly abundant (9.45 $\times 10^{-9}$). 

Recently Sivaraman et al. (2012), performed an experiment to obtain the IR spectra
of the formamide in the ice phase.
They used experimental setup based at The Open University, UK (Sivaraman et al., 2008) 
to simulate astrochemical ices and their irradiation environments. The instrument was
operated at base pressure of the order of $10^{-10}$Torr, and it could go down up to
the temperature 28K. Low temperature was achieved by using a closed cycle helium 
cryostat. A CaF$_2$ substrate was placed at the end of the cryostat onto 
which the molecular gases were directly deposited to form multilayer targets. 
Sample temperature measurements were carried out using a silicon (Si) diode 
sensor calibrated using the calibration curve provided by the Scientific Instruments.
Formamide samples 99.5 \% pure (from Sigma Aldrich), were used. Before introducing the 
formamide vapour into the chamber, the liquid sample was processed by 
three freeze-pump-thaw cycles to degas any absorbed impurities.  
The sample was then allowed to return to room temperature before extracting 
the vapours to form the ice on the CaF$_2$ substrate.

\begin{table*}
\scriptsize{
\centering
\hskip -2.5cm
\vbox{
\addtolength{\tabcolsep}{-6pt}
\caption{Vibrational frequencies of Formamide in gas phase, H$_2$O ice and methanol containing grains at B3LYP/6-311G++** level of theory}
\begin{tabular}{|c|c|c|c|c|c|c|c|c|c|}
\hline
{\bf Species}&{\bf Peak positions }&{\bf Absorbance}&{\bf Peak positions }&{\bf Absorbance}&{\bf Peak positions }&{\bf Absorbance}&{\bf Peak position}&{\bf Absorbance}&{\bf Peak positions}\\
{}&{(Gas phase)}&{}&{(H$_2$O ice)}&{}&{(Methanol ice) }&{}&{(Mixed ice)}&{}&{by Experiment}\\
&(Wavenumber in cm$^{-1}$)&{}&(Wavenumber in cm$^{-1}$)&{}&(Wavenumber in cm$^{-1}$)&{}&(Wavenumber in cm$^{-1}$)&{}&(Wavenumber in cm$^{-1}$)\\
\hline
&535.27&217.71&574.75&24.36&574.72&23.96&574.72&24.34&-\\
&566.2&11.04&600.39&202.88&598.11&297.18&598.25&305.54&-\\
&656.9&134.94&668.7&202.88&669.25&200.41&670.22&198.46&-\\
&1035.82&2.7188&1053.77&5.21&1053.25&5.2&1055.15&5.18&1022\\
&1067.82&3.46&1082.36&2.24&1082.29&2&1083.01&2.16&1056\\
&-&-&-&-&-&-&-&-&1117\\
&-&-&-&-&-&-&-&-&1172\\
&1280.07&115.04&1309.74&134.26&1308.97&135.88&1309.8&134.71&1226\\
&-&-&-&-&-&-&-&-&1334\\
{\bf H$_2$NCHO}&1416.33&27.68&1387.4&168.36&1388.86&158.32&1388.84&167.9&1386\\
&1654.2&44.98&1602.01&311.72&1604.49&300.78&1603.19&293.25&1628\\
&1697.36&405.23&1641.11&504.78&1643.25&503.21&1640.75&523.72&1698\\
&3000.13&88.48&2996.46&32.04&2997.12&33.95&2991.10&32.51&2895\\
&-&-&-&-&-&-&-&-&3179\\
&3569.63&27.22&3398.15&183.5&3405.10&175.06&3396.87&182.07&3372\\
&3715.58&27.94&3528.69&156.17&3536.57&149.65&3529.07&157.64&-\\
\hline
\end{tabular}
}}
\end{table*}

\begin {figure}
\vskip 0.6cm
\centering{
\includegraphics[height=10cm,width=10cm]{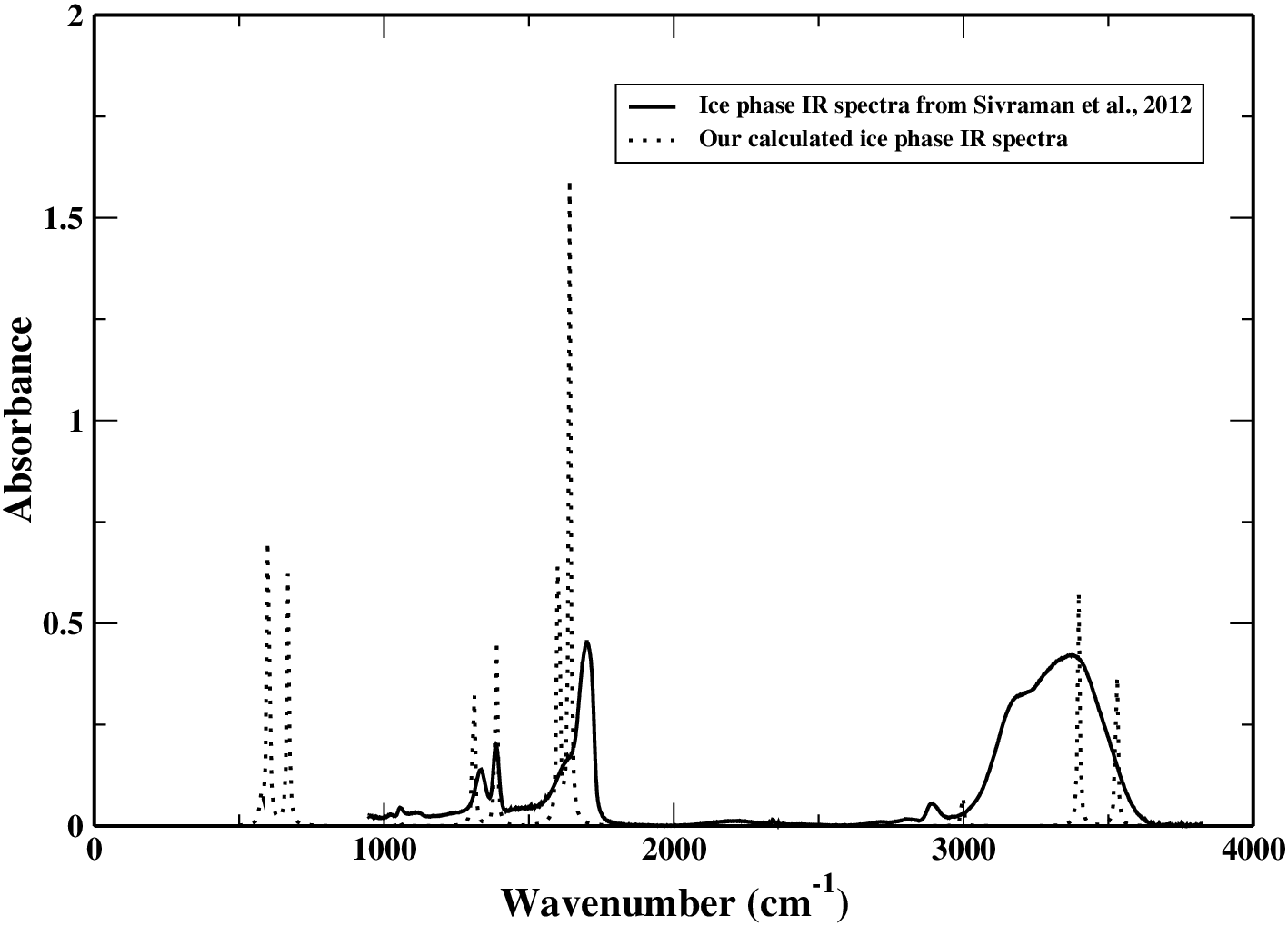}
}
\caption{Comparison between the ice phase formamide IR spectra obtained by our quantum chemical 
approach and experimental approach by Sivaraman et al., 2012}
\label{fig-7}
\end {figure}
We compare our theoretical result with this recent experiment.
Figure 7 shows the normalized infrared spectra of formamide molecule in the ice phase  
(where H$_2$O was used as a solvent) which we calculate
along with the experimentally obtained infrared spectra of formamide at 30K before irradiation.
Peak positions are given in Table 3. It is clear from Table 3 that some of the calculated 
peaks are very close to the experimental values. 
For example, our calculated peak position is at 1387.40 cm$^{-1}$, whereas the experimentally
obtained peak location is at 1386 cm$^{-1}$. Beside this, there are a few more peaks which 
are close to the experimental values (Table 3). 
{\bf To have an idea about the effect of solvent upon the spectrum, we considered
different kind of ice as a solvent. In general, we have considered the water ice but depending
upon the properties around the molecular cloud, 
ice composition may be different (Das et al., 2010). Keeping this, first 
we consider pure methanol ice instead of pure water ice and study the changes in the
peak positions and intensities (Table 3). Second, based on the observational results,
we consider a mixed ice, which consists of 70\% water, 20\% methanol and 10\% CO$_2$
molecules and noted down the spectral properties in Table 3.
From Table 3, it is evident that the peak positions of formamide in methanol containing 
ice and mixed ice are slightly shifted in compare to the formamide in pure water ice.} 
Though some peak positions are closely coinciding with our theoretical calculations, some
observed peaks are well above our values. Our quantum chemical calculations are based upon the
Ab initio methods in GAUSSIAN 09, which employ the Born-Openhiemer approximation in generating the energy
expressions. It then allows us to separate the nuclear and electronic degrees of freedom. The energy is the 
electronic energy parameterized by the frozen locations of the nuclei and because they are frozen, the model
simulates at 0K, whereas the experiment was performed at 30K. Moreover, here we are not
considering the clusters of formamide for calculating the spectra. Rather, we are including 
one formamide molecule in a spherical cavity, which is immersed in a continuous medium
with a dielectric constant. Here the solute-solvent electrostatic interactions are treated at the dipole level.
The solvent effect brings significant changes in the geometrical parameters of formamide. 
Our model confirms that the polarization of
the solute by the continuum has important effects on the absolute and
relative solvation energies, which, in turn, shift the frequency and
vary the intensity in the infrared mode.
On the other hand, during the experiment, several clusters could be formed. 
So, an ideal match is not expected.

\begin{table*}
\addtolength{\tabcolsep}{-4pt}
\centering
\scriptsize{
\hskip -4.5cm
\vbox{
\caption{Electronic transitions of Formamide at B3LYP/6-311++G** level theory   in gas phase and H$_2$O ice}
\begin{tabular}{|c|c|c|c|c|c|c|c|c|c|c|}
\hline
{\bf Species}&{\bf Wavelength}&{\bf Absorbance}&{\bf Oscillator strength }&{\bf Transitions}&{\bf Contribution}&{\bf Wave length}&{\bf Absorbance}&{\bf Oscillator strength}&{\bf Transitions}&{\bf Contribution}\\
{}&{(gas phase)}&{}&{}&{}&{(\%)}&{(H$_2$O ice)}&{}&{}&{}&{(\%)}\\
&(in nm)&&&&&(in nm)&&&&\\
\hline
&184.8&3365&0.0736&H-0$\rightarrow$ L+2&99&166.7&17903&0.189&H-0$\rightarrow$ L+2&43\\
&163.4&13685&0.3215&H-1$\rightarrow$ L+1&65&114.5&12348&0.088&H-3$\rightarrow$ L+1&56\\
&122.5&5963&0.0122&H-3$\rightarrow$ L+0&98&92.2&13377&0.0236&H-0$\rightarrow$ L+16&93\\
{\bf H$_2$NCHO}&115.5&5647&0.0531&H-2$\rightarrow$ L+1&33&-&-&-&-&-\\
&100.1&11453&0.1258&H-3$\rightarrow$ L+5&55&-&-&-&-&-\\
&92.0&11963&0.0326&H-0$\rightarrow$ L+16&92&-&-&-&-&-\\
\hline
\end{tabular}}
}
\end{table*}
Electronic absorption spectral parameters of formamide molecule in gas phase as well as in the ice 
phase are highlighted in Table 4. An electronic absorption spectrum of formamide molecule in gas 
phase is characterized by 3 intense peaks. These three transitions 
occurred at 163.4, 100.1, 92.0 nm with  major contributions from H-1$\rightarrow$ L+1, 
H-3$\rightarrow$ L+5, 
H-0$\rightarrow$ L+16 transitions respectively. With these strong peaks, some substructures are also prominent in the gas phase, which are completely absent in the ice phase. As in the other cases, these intense peaks 
are due to the transition from HOMO to LUMO (Table 4). 

\section{Conclusion}
In this paper, we have explored the infrared and electronic absorption spectra of 
various complex molecules, which could be treated as the precursors of 
adenine, glycine and alanine in several astronomical situations.
{\bf Since we use a realistic scenario with hydro-chemical evolutions 
with accurate rate coefficients (Majumdar et al. 2012) our results could be 
significant for future observations.}

In comparison with the gas phase spectra, the spectral properties in the ice phase
are significantly different. Various peak positions are shifted and some peaks
completely disappeared. Some substructures became pronounced also in the ice phase.
{\bf Based on the observational results, we simulated our grain mantle differently which have
different ice compositions. We considered the ice to be that of pure water or 
pure Methanol or even mixed (70\% water 20\% methanol and 10\% CO$_2$) in nature.
We have highlighted the dependence of the spectral changes on the solvent types}

We computed the chemical evolution of adenine, alanine \& glycine along with their precursor molecules.
We pointed out that the computed abundances of the precursor molecules are within the observational limits
and could be detected in future. Based on the abundances of these precursor molecules, one could
estimate the abundances of the molecules of our interest. We made a comparison between our calculated 
infrared spectra of formamide and experimentally obtained infrared spectra. 
Some of the peak locations are found to coincide while a few others
do not match. We presented possible reasons behind this.

There are ample debates on  whether alanine, glycine or adenine has been seen in interstellar medium or not, 
It is possible that the abundances of these molecules are very small and the present day equipments lack 
sufficient sensitivities. However, our paper probes the feasibility of observing the precursors of these
very important molecules and computed abundances seem to indicate that they are observable, in principle. 
We have also computed the spectral features, i.e., the wavelengths 
where the characteristic peaks would be observed. We 
anticipate that these results would be appreciated by observers.

\section{Acknowledgment}
We acknowledge the help of Dr. B. Sivaraman, INSPIRE Faculty (IFA-11CH -11), 
Indian Institute of Science, B. N. Raja Sekhar of AMPD, BARC at RRCAT Indore and N. J. 
Mason of The Open University for supplying us with the experimental IR data for Formamide.
LM, SKC \& SC are grateful to DST for the ﬁnancial support through a project 
(Grant No. SR/S2/HEP-40/2008) and AD thanks ISRO respond project (Grant No. ISRO/RES/2/372/11-12).

\end{document}